\documentclass[manuscript]{acmart}
\AtBeginDocument{%
  \providecommand\BibTeX{{%
    \normalfont B\kern-0.5em{\scshape i\kern-0.25em b}\kern-0.8em\TeX}}}

\setcopyright{acmcopyright}
\copyrightyear{2023}
\acmYear{2023}

\usepackage{multirow} 

\copyrightyear{2023}
\acmYear{2023}
\setcopyright{acmcopyright}\acmConference[LAK 2023]{LAK23: 13th International Learning Analytics and Knowledge Conference}{March 13--17, 2023}{Arlington, TX, USA}
\acmBooktitle{LAK23: 13th International Learning Analytics and Knowledge Conference (LAK 2023), March 13--17, 2023, Arlington, TX, USA} \acmPrice{15.00}
\acmDOI{10.1145/3576050.3576147} \acmISBN{978-1-4503-9865-7/23/03}

\begin{document}

\title{Trusting the Explainers: Teacher Validation of Explainable Artificial Intelligence for Course Design}

\author{Vinitra Swamy}
\email{vinitra.swamy@epfl.ch}
\affiliation{%
  \institution{EPFL}
  \country{Switzerland}
}

\author{Sijia Du}
\email{sijia.du@epfl.ch}
\affiliation{%
  \institution{EPFL}
  \country{Switzerland}}

\author{Mirko Marras}
\email{mirko.marras@acm.org}
\affiliation{%
  \institution{University of Cagliari}
  \country{Italy}}

\author{Tanja Käser}
\email{tanja.kaeser@epfl.ch}
\affiliation{%
  \institution{EPFL}
  \country{Switzerland}
}

\renewcommand{\shortauthors}{Swamy et al.}

\begin{abstract}

Deep learning models for learning analytics have become increasingly popular over the last few years; however, these approaches are still not widely adopted in real-world settings, likely due to a lack of trust and transparency. In this paper, we tackle this issue by implementing explainable AI methods for black-box neural networks. This work focuses on the context of online and blended learning and the use case of student success prediction models. We use a pairwise study design, enabling us to investigate controlled differences between pairs of courses. Our analyses cover five course pairs that differ in one educationally relevant aspect and two popular instance-based explainable AI methods (LIME and SHAP). We quantitatively compare the distances between the explanations across courses and methods.
We then validate the explanations of LIME and SHAP with $26$ semi-structured interviews of university-level educators regarding which features they believe contribute most to student success, which explanations they trust most, and how they could transform these insights into actionable course design decisions. Our results show that quantitatively, explainers significantly disagree with each other about what is important, and qualitatively, experts themselves do not agree on which explanations are most trustworthy. All code, extended results, and the interview protocol are provided at \href{https://github.com/epfl-ml4ed/trusting-explainers}{https://github.com/epfl-ml4ed/trusting-explainers}.
\end{abstract}



\maketitle

\section{Introduction}

There is a compelling need for interpretability in models dealing with human data, especially in education. \citet{conati2018ai} strongly argues for interpretable models in education, especially in settings where students can see the effect of a decision but not the reasoning behind it (e.g., open learner modeling). Furthermore, \citet{DBLP:conf/lak/NazaretskyCA22} has found that transparency is an essential factor in increasing educators' trust in AI-based educational technology.

The explainable machine learning for education community is only now emerging with few works over the last two years, most of which focus on implementing only one interpretability method for a downstream educational task, using either LIME or SHAP (e.g., \cite{hasib2022lime, baranyi2020interpretable, scheers2021interactive, mu2020towards, pei2021}). Therefore, there exists a pervasive research gap in contrasting multiple explainable AI (XAI) methods to \textit{validate} which methods are more suited or trustworthy for human-centric tasks. In prior work, \citet{Swamy2022} examined and compared five popular instance-based explainability methods on student success prediction models for five different massive open online courses (MOOCs). The experiments from this paper show that the feature importance distributions extracted by different explainability methods for the same model and underlying students differ significantly from each other. \citet{marras2021can} focused on how the feature importance differs across different course modalities (MOOCs and flipped classrooms) using random forest models. Their work demonstrates that the ensemble of optimal features varies depending on the the course design and the characteristics of the population. Inspired by these papers, we are interested in how the explanations of success prediction models from courses with contrasting educational aspects (in terms of instructional design and target audience) vary from each other. We aim to extend \citet{Swamy2022}'s result, going beyond ``are the explanations generated by the methods systematically different?" towards ``which method is most valid for the downstream application?". To estimate this, we obtain the closest approximation of ground truth through semi-structured expert interviews with educators, aiming to understand which XAI methods the experts trust most and how these insights can be actionable for course design.

Specifically, we select pairs of courses with different educational factors: four pairs with differences in instructional design in terms of setting (MOOC vs. flipped classrooms), active learning (high vs. low), and optimality of design for the target audience (high vs. low passing rate) and one pair with differences in student population, measured by the instruction language (English vs. French). While learning science literature points to many other significant factors influencing students' learning behavior and success in online and blended courses, we limited ourselves to the factors measurable in our dataset based on the information about the curriculum and demographics accessible to us. We first train models of student success prediction separately for each course using the bidirectional LSTM (BiLSTM) network proposed by \citet{Swamy2022, swamy22b}. We then apply two popular instance-based explainability methods \cite{lime, shap} to each model. We formulate comparable feature importance scores for each explainer. To contrast the feature importance distributions quantitatively, we use different distance measures comparing feature ranks (\textit{Spearman’s Rank-Order Correlation}) and similarity of the top ten features (\textit{Jaccard Similarity}).
For two course pairs (setting and active learning), we conduct expert interviews with $26$ STEM professors to build validation and trust in the generated explanations. Participants are asked to rank the features they find important for differentiating passing and failing student behavior, identify the explanations they trust the most (with the underlying method obfuscated), and brainstorm actionable course design changes from these insights. Specifically, in our study, we aim to answer the following research questions:
\begin{enumerate}
\small
    \item How similar are the explanations of different explainability methods for a specific course?
    \item Do explanations capture a controlled difference across pairs of courses with different settings?
    \item Which explainability methods do domain experts trust most?
    \item Can experts use XAI insights for actionable course design decisions?
\end{enumerate}

We provide results from a nine-course quantitative analysis and a semi-structured expert interview study regarding the trustworthiness of the LIME and SHAP explainers. Our quantitative results show that the explainers do not agree with each other about important features on any of the individual course models. From a qualitative lens, our expert interviews reveal that educators do not consistently prefer any one explainer and often even choose a confounder explanation. Although our study reveals a lack of trust in explainer agreement, over $85\%$ of educators are able to use these insights to generate actionable decisions with the goal of improving student learning outcomes.

\section{Methodology}
\label{sec:method}

\begin{figure*} 
     \centering
     \includegraphics[width=\linewidth]{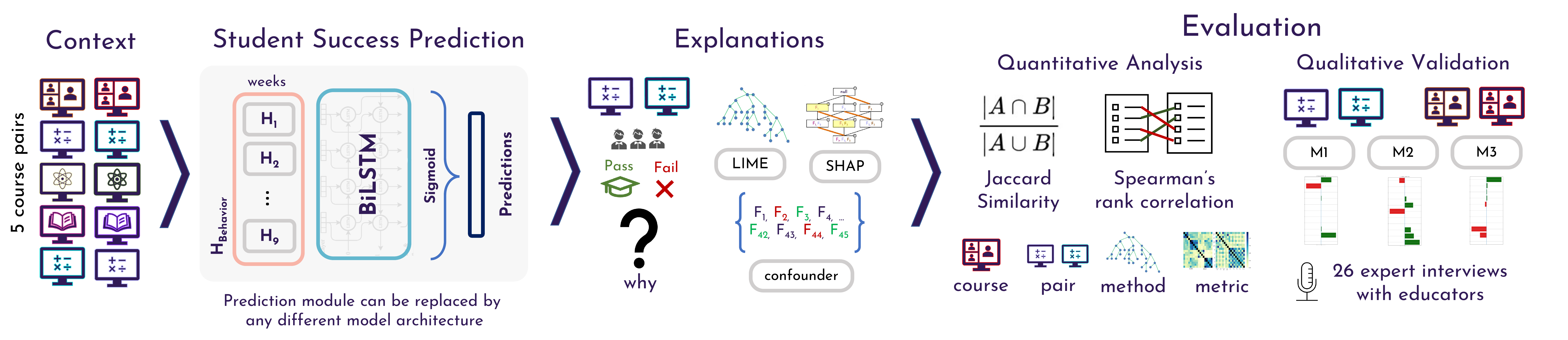}
     \caption{Our framework involves four steps: 1) extracting clickstream data from five pairs of courses, 2) modeling course data for success prediction, 3) using XAI methods to extract feature importance scores, 4) quantitative and qualitative evaluation of explainers.}
     \label{fig:pipeline}
     \vspace{-4mm}
 \end{figure*}

With our methodology, we aim to quantitatively and qualitatively compare and validate the explanations from success prediction models across different settings and assess whether they can build the basis for actionable interventions. Fig. \ref{fig:pipeline} illustrates the main steps of our XAI evaluation framework. While explanations of model predictions are essential to increase trust \cite{DBLP:conf/lak/NazaretskyCA22} and enable evidence-based interventions in any educational scenario, we focus on online and blended learning scenarios, aiming to predict student success. In a first step, we extract pairs of courses $(c, c')$ with `controlled' differences based on literature influencing students' behavior and learning experience. For example, courses $c$ and $c'$ might both be Bachelor's level mathematics courses, but exhibit a different instructional design. Alternatively, they might have the same instructional design but a different target audience. This study design enables us to hypothesize about potential causes for a success/failure and therefore validate the provided explanations. In a second step, we extract the interaction logs for all selected courses and train a success prediction model separately for each course. Next, we apply post-hoc explainability methods to retrieve explanations for a single student. The two final steps then consist of the quantitative evaluation and the semi-structured interviews with experts. 

\subsection{Learning Context}
Online and blended learning are becoming immensely popular and received a further boost in popularity due to the COVID-19 pandemic \cite{impey2021moocs}. Unfortunately, MOOCs suffer from a high drop-out and low success rates. Furthermore, prior research has shown that students often do not possess the self-regulated learning skills necessary to succeed in flipped classrooms (a form of blended learning)\cite{hirsto2019exploring}. Both MOOC and flipped classroom scenarios could therefore benefit from models being able to predict student success \textit{accurately} and in an \textit{interpretable} way. In the scope of this study, we therefore focus on the context of MOOCs and flipped classroom courses. 

\vspace{1mm} \noindent \textbf{Study Design}. 
\label{sec:study-design}
We opted for a pairwise study design as a basis for the validation of explanations. Specifically, we extracted pairs of courses $(c,c')$ with opposite (or largely different) characteristics, which are known to have an impact on student behavior and learning outcomes. We assumed that these behavioral differences would be captured by the student success prediction model and therefore reflected in the related explanations.
From previous research, two major factors influencing learning in online settings have emerged: \textit{population} and \textit{instructional design}. 

Multiple studies (e.g., \cite{ke2013online}) have reported the important influence that demographics in terms of age, culture, language, and gender can have when designing online learning. For example, it has been shown that help-seeking behavior differs across cultures \cite{Ogan15} and that the preference for specific learning activities is influenced by culture and socio-economic background \cite{Rizvi22}. A recent large-scale analysis across multiple MOOC providers has confirmed the impact of the language of instruction, cultural background, or localized course design on learners' educational experiences \cite{ruiperez22}. Furthermore, \citet{Kiziclec13} has also found differences in engagement across age groups and gender. Unfortunately, the data available to us contained only sparse demographic information, with gender and geographic location only available for very small subsets of course participants ($<9\%$ for most of the courses). We therefore used the instruction language as a proxy of the target audience. Table \ref{tab:study_design} lists the according study design for the factor \textit{population}.

A large body of prior work (e.g., \cite{GASEVIC201668,JUNG2019377,Yang17}) has also demonstrated the impact of the instructional design on students' learning experience and success. For example, \citet{JUNG2019377} has shown that instructional design factors are stronger indicators of a successful MOOC experience than the content of the course or demographic factors. \citet{Yang17} has found that the quality of a course influences the persistence of students in MOOCs. These findings have been confirmed by \citet{GASEVIC201668}, who demonstrated that the accuracy of predictive models can be increased by including indicators that describe the instructional design. In our study, we focused on three dimensions of \textit{instructional design} that we could measure based on the available data. The first dimension is active learning, which was found to be an important factor for engagement and popularity as well as learning outcomes in MOOCs \cite{Cagiltay20,Hew16,Koedinger15, mejia22}. In our case, we approximate the amount of active learning in a course by calculating the average weekly quiz ratio, i.e. the number of weekly quizzes divided by the number of weekly videos. The second investigated dimension is the setting of the course (MOOC or flipped classroom). \citet{marras2021can} compared a large number of learning indicators extracted from previous literature in flipped and online learning settings and found differences in feature importance across settings. For our study, we therefore compared pairs of courses that have been taught as a MOOC as well as in a flipped setting. The last dimension that we included is the optimality of the course design. Prior research has demonstrated that the course design should be adapted to the target population \cite{Kiziclec13,Rizvi22}. For our context, we approximated the optimality of the course design by the course success rate. We chose a pair for comparison, where the same course (in terms of content, level, and language) was taught in two iterations, with a substantial higher success rate in the second iteration. 
The three dimensions for \textit{instructional design} along with the related indicators and selected course pairs are listed in Table \ref{tab:study_design}.

\begin{table*}[!htb]
\small
\caption{\textbf{Study Design}. We performed pairwise comparisons across nine courses within two factors influencing interaction behavior (instructional design and population). We analyze three dimensions within course design and one dimension for population.} 
\vspace{-1mm}
\resizebox{\textwidth}{!}{
\begin{tabular}{l||cccccccc|cc}
\toprule
\textbf{Factor} &
\multicolumn{8}{c|}{\textbf{Course Design}} &
  \multicolumn{2}{c}{\textbf{Population}} \\ 
  \midrule
  \textbf{Dimension} &
\multicolumn{4}{c|}{\textbf{Setting}} &
  \multicolumn{2}{c|}{\textbf{Active Learning}} &
  \multicolumn{2}{c|}{\textbf{Optimality}} &
  \multicolumn{2}{c}{\textbf{Language}} \\
  \midrule
  \textbf{Indicator} &
  \multicolumn{2}{c}{Flipped} &
  \multicolumn{2}{c|}{MOOC} &
  \begin{tabular}[c]{@{}c@{}}High quiz to \\ video ratio\end{tabular} &
  \multicolumn{1}{c|}{\begin{tabular}[c]{@{}c@{}}Low quiz to \\ video ratio\end{tabular}} &
  \begin{tabular}[c]{@{}c@{}}Higher \\ success rate\end{tabular} &
  \begin{tabular}[c]{@{}c@{}}Lower \\ success rate\end{tabular} &
  English &
  French \\
  \midrule
  \textbf{Course ID} &
  Flip LA &
  Flip FP &
  MOOC LA &
  \multicolumn{1}{c|}{MOOC FP} &
  AN$_2$ &
  \multicolumn{1}{c|}{Geo} &
  AN$_2$ &
  AN$_1$ &
  VA$_1$ &
  VA$_2$ \\ 
 \bottomrule
\end{tabular}}
\label{tab:study_design}
\end{table*}

\vspace{1mm} \noindent \textbf{Data Collection}.
Our experiments are based on interaction and metadata collected from seven MOOCs and two flipped classroom courses \cite{hardebolle2022gender} of  
École Polytechnique Fédérale de Lausanne, recorded across three online platforms (Coursera, EdX, and Courseware).  The MOOCs were accessible without any restrictions, i.e. worldwide and at any time. All the selected courses organized the content in weeks. Each week included video lectures introducing the important concepts as well as quizzes for self-assessment. Furthermore, students were requested to also take graded assignments (scored up to $100$ points), which served as a basis for obtaining the course certificate. The number of assignments and the minimal score needed for passing the MOOC was up to the instructor. While the flipped courses were recorded on the same platform, they offered restricted access only. Both of the flipped courses were compulsory (in the Bachelor's and Master's, respectively). Students of the class were required to perform online pre-class activities including watching lecture videos and solving associated quizzes.
Table \ref{tab:courses} lists detailed information about all the included courses. The data collection and analysis was approved by the university’s ethics review board (HREC 058-2020/10.09.2020, 096-2020/09.04.2020).

The log data for both MOOCs and flipped courses was collected in the form of a time-wise clickstream per student. Specifically, students enrolled in a course $c$ interact with the learning objects $\mathbb{O}^c$ from that course. In our study, we assume that the learning objects can be either videos or quizzes, but the framework can be easily extended to other types (e.g., forum posts, textbook readings). All learning objects $\mathbb{O}^c$ from a course $c$ are associated with a specific course week $w$. The extraction of this weekly course schedule allows engineered learning indicators specific to the course design (see Section \ref{sec:feat_extract}). We denote the \emph{interactions} of a student $s$ in a course $c$ as a time series $I_s^c = \{i_1, \ldots, i_K\}$ (e.g., a sequence of video plays and pauses, quiz submissions). We assumed that each interaction $i$ is represented by a tuple $(t, a, o)$, including a \emph{timestamp} $t$, an \emph{action} $a$ (videos: load, play, pause, stop, seek, speed; quiz: submit), and a \emph{learning object} $o \in \mathbb{O}^c$ (video, quiz). We finally denote the \emph{binary success label} (pass-fail) for student $s$ in course $c$ as $l_{s,c}$. 

\begin{table*}[!htb]
\caption{\textbf{Learning Context}. Detailed instructional design and student population information on the nine selected courses for this XAI study. The relevant indicators for the pairwise selection are denoted in bold.}
\label{tab:courses}
\vspace{-2mm}
\small
\resizebox{\textwidth}{!}{
\begin{tabular}{llllllrrrr}
\toprule
\textbf{Course Title} & \textbf{Course ID} & \textbf{Setting} & \textbf{Field$^1$} &  \textbf{Level} & \textbf{Language}  & \multicolumn{1}{r}{\textbf{\begin{tabular}[c]{@{}c@{}} Weeks\end{tabular}}} & 
\multicolumn{1}{r}{\textbf{\begin{tabular}[c]{@{}c@{}} Quiz/Video$^2$ \end{tabular}}} & 
\multicolumn{1}{r}{\textbf{\begin{tabular}[c]{@{}c@{}} Students$^3$\end{tabular}}} & \multicolumn{1}{r}{\textbf{\begin{tabular}[c]{@{}c@{}} Success Rate \end{tabular}}} \\
\midrule
Algebra (part 2)& MOOC LA & \textbf{MOOC} & Math & Prop & French & 4 & 2.13 & 170 & $0.67$\\
Algebre lineaire&Flip LA&\textbf{Flipped} & Math & BSc & French &10 & 1.76 & 214 &$0.59$\\
Functional Programming Principles in Scala&MOOC FP&	\textbf{MOOC}& CS& BSc&	English&	6& 0.52 &	3565& $0.48$\\
Functional programming&	Flip FP& \textbf{Flipped}&CS&MSc&	English&	10& 0.21&	218& $0.62$\\
African Cities - An introduction to urban planning&VA$_1$&MOOC& SS&	BSc&	\textbf{English}&	12&	5.42 & 5643& $0.10$\\
African Cities - An introduction to urban planning&VA$_2$& MOOC& SS& Prop&\textbf{French}&12&5.42&4699& $0.05$\\
Analyse Numérique pour Ingénieurs&	AN$_1$& MOOC& Math&	BSc&	French&	9&9.14&	506& $\mathbf{0.08}$\\
Analyse Numérique pour Ingénieurs&	AN$_2$&MOOC&Math&BSc&	French&	9&\textbf{9.14}&	506& $\mathbf{0.71}$\\
Éléments de Géomatique&	Geo& MOOC& Eng.& BSc&French&11&\textbf{3.89}&452&$0.45$ \\
\bottomrule
\end{tabular}}
\scriptsize{$^1$\textit{CS}: Computer Science; \textit{Math}: Mathematics; \textit{SS}: Social Sciences; \textit{Eng.}: Engineering.} \; \; 
\scriptsize{$^2$Nr. of weekly quizzes/Nr. of weekly videos, averaged over all course weeks.} \; \; \\
\scriptsize{$^3$For MOOCs, after removing early-dropout students according to \cite{swamy22b}.}
\end{table*}

\subsection{Student Success Prediction}
\label{sec:blackbox-model}
The collected student log data in the form of time-series of interactions builds the basis for extracting features serving as indicators of learning. The engineered features are fed into the success prediction model to obtain a pass-fail prediction $l_{s,c}$ for each student $s$ in course $c$, which can then be interpreted and explained.

\vspace{1mm} \noindent \textbf{Features}. \label{sec:feat_extract} A large body of work has focused on developing success prediction models for MOOCs (e.g., \cite{gardner2018student}) and flipped classroom courses (e.g., \cite{lee2022affects}). A multitude of feature sets and classifiers tied to a specific course or a small set of similar courses have been proposed. In their recent meta-analysis, \citet{marras2021can} performed a literature search to retrieve prior research proposing feature sets for success prediction models in online settings. After removal of overlapping sets, the combined feature set contains over $100$ features, which we use as input for our models. Each feature is extracted per student $s$ and week $w$ in course $c$, meaning that each feature is a time-series of length $W$ (where $W$ is the number of weeks of the course). The combined feature set has been retrieved from nine different papers \cite{marras2021can, DBLP:conf/edm/AkpinarRA20,boroujeni2016quantify,chen2020utilizing,lalle2020data,DBLP:journals/eait/LemayD20,DBLP:conf/aied/MbouzaoDS20,DBLP:journals/eait/MubarakCA21,DBLP:journals/tlt/WanLYG19}. Given the high number of features (110), we will give a coarse overview and categorization of the features only and refer to the original work as well as to the public GitHub repository\footnote{https://github.com/epfl-ml4ed/flipped-classroom} of \citet{marras2021can} for a detailed documentation of the features. The subset of features identified as important by the explainers and presented to educators is showcased in Table \ref{tab:features}.


The extracted features can be roughly divided into two categories: engineered features that serve as indicators of students' self-regulated learning (SRL) behavior and raw patterns extracted from the clickstream data logs. For engineered features, we have features with dimensions of \textit{effort} (effort regulation), \textit{regularity} (time management), \textit{proactivity} (time management), and \textit{control} (metacognition), which can all be associated with success in online higher education \cite{broadbent2015self}.
Features in the \textit{\textbf{Effort}} dimension \cite{chen2020utilizing,DBLP:journals/eait/LemayD20,lalle2020data,DBLP:conf/aied/MbouzaoDS20,DBLP:journals/tlt/WanLYG19, DBLP:conf/edm/AkpinarRA20, he2018measuring} monitor the extent to which a student is engaged in the course, which has been shown to be essential for success (e.g., \cite{cho2013self}).
Features in the \textit{\textbf{Regularity}} dimension \cite{boroujeni2016quantify, he2018measuring} monitor the extent to which the student follows a regular study pattern. A regular working schedule has been proven to be predictive of success in MOOCs \cite{boroujeni2016quantify} and flipped settings \cite{DBLP:journals/ce/JovanovicMGDP19}. 
Features in the \textit{\textbf{Proactivity}} dimension \cite{marras2021can, boroujeni2016quantify,DBLP:journals/tlt/WanLYG19} attempt to measure how much students are ahead or behind the schedule of the course. Proactivity has been shown to play an important role in student success \cite{geertshuis2014preparing, boroujeni2016quantify}. 
The forth dimension of engineered features, \textit{\textbf{Control}} \cite{lalle2020data,DBLP:journals/eait/LemayD20,DBLP:journals/eait/MubarakCA21}, includes indicators of fine-grained video behavior, that constitute a proxy for assessing to what extent students can control the cognitive load of the video lectures. Prior research has shown that the flow of videos in MOOCs can lead to cognitive overload \cite{biard2018effects}.
For patterns stemming directly from the raw interaction logs (notably not engineered indicators), \citet{DBLP:conf/edm/AkpinarRA20} extracted consecutive \textit{\textbf{Subsequences}} of $n$ clicks extracted from the session clickstreams of a blended course. The resulting feature set is consequently huge  (e.g., $350$ features for course Flip LA). We therefore only included the sub-sequence features found important in \citet{marras2021can}, resulting in $110$ total features.

\begin{table*}[!htb]
\centering
\small
\caption{\textbf{Features}. For brevity, we only list the $22$ features that have been identified as important by at least one explainability method in our analysis in Section \ref{sec:results}. Remaining features can be found in \citet{marras2021can}.}
\label{tab:features}
\vspace{-2mm}
\resizebox{\textwidth}{!}{
\begin{tabular}{@{}llll@{}}
    \toprule
    \textbf{Dimension} & \textbf{Feature (simplified names)} &  & \textbf{Description} \\ \midrule
    \multirow{5}{*}{\textit{Effort}} & check-check-check-quiz & \cite{DBLP:conf/edm/AkpinarRA20} & The amount of times the student checks problems three times in a row. \\
    & correct-time-quiz & \cite{DBLP:journals/tlt/WanLYG19} & Total time spent divided by the number of correct problems. \\
    & distinct-probs-quiz & \cite{DBLP:journals/tlt/WanLYG19} & The total number of distinct problems attempted by the student. \\
    & num-submit-quiz & \cite{DBLP:journals/tlt/WanLYG19} & The number of submissions performed for a quiz, on average. \\
    & total-time-vid & \cite{DBLP:conf/edm/AkpinarRA20} & The total (cumulative) time that a student has spent on video events. \\
     \midrule
    \multirow{5}{*}{\textit{Regularity}} 
    & active-participation-weekly-vid & \cite{marras2021can} & The number of videos the student watched fully over total loaded videos (per week). \\ 
    & attendance-rate & \cite{he2018measuring} & The number of videos played over the total number of videos released. \\
    & hourly-freq-regular & \cite{boroujeni2016quantify} & The extent to which the hourly pattern of user’s activities is repeating over days. \\
    & watch-ratio-vid & \cite{he2018measuring} & Ratio of amount of video watched for videos a student opens. \\
    & std-time-session & \cite{chen2020utilizing} & The standard deviation of time spent from a login to the end of the session. \\
     \midrule
    \multirow{5}{*}{\textit{Proactivity}}
    & eager-view-vid & \cite{marras2021can} & The extent to which the student approaches a video early.\\
    & timely-view-vid & \cite{marras2021can} & The extent to which the student approaches a video in the right week.\\
    & eager-view-quiz & \cite{marras2021can} & The extent to which the student approaches a quiz early.\\
    & ratio-clicks-weekend & \cite{chen2020utilizing} & The ratio between clicks in weekdays and weekends. \\
    & std-correct-time-quiz & \cite{DBLP:journals/tlt/WanLYG19} & Variance of total time spent divided by the number of correct problems. \\
     \midrule
     \multirow{7}{*}{\textit{Control}} 
     & avg-len-seek-vid & \cite{lalle2020data} & The student's average seek length (seconds). \\
    & freq-pause-vid & \cite{lalle2020data} & The frequency between every Video.Pause action and the following action. \\
    & freq-play-vid & \cite{lalle2020data} & The frequency of the play event in the students’ sessions. \\
    & play-stop-play-vid & \cite{DBLP:conf/edm/AkpinarRA20} & The amount of times the student plays a video, stops, and plays another one. \\
    & play-pause-load-vid & \cite{DBLP:conf/edm/AkpinarRA20} & The amount of times the student plays a video, pauses, and plays again. \\
    & pause-speedchange-play-vid & \cite{DBLP:conf/edm/AkpinarRA20} & The amount of times the student pauses, changes the speed, and plays the video. \\
    & speed-vid & \cite{DBLP:journals/eait/LemayD20} & The average speed the student used to play back a video. \\
     \bottomrule
    \end{tabular}}
\end{table*}
\renewcommand{\arraystretch}{1.0}

\vspace{1mm} \noindent \textbf{Predictive Model}.
To predict the pass-fail label $l_{s,c}$ of student $s$ in course $c$ based on the extracted behavioral features, we relied on the neural architecture based on \emph{Bidirectional LSTMs} suggested by \cite{Swamy2022}. The model input is represented by $H$, i.e., the extracted behavior features, having a shape of $|S| \times W \times 110$, where $|S|$ and $W$ denote the number of students and weeks, respectively. NaN values were replaced with the minimum value of each respective feature. The architecture is composed of two \emph{BiLSTM} layers of size $32$ and $64$ (using a loopback of $3$) and a \emph{Dense} layer (with Sigmoid activation) having a hidden size of $1$. The model outputs the probability $p_{s,c}$ that the student $s$ will pass course $c$.

\subsection{Generating Explanations}
\label{sec:explanabilitymethods}
While deep learning models have shown highly accurate performance over traditional machine learning methods, this accuracy comes at the cost of interpretability. Neural networks are therefore referred to as black-box models, emphasizing their lack of transparency. Post-hoc explainability methods can be applied to the predictions of a black-box model in order to gain interpretablity in the form of feature importance score. We used LIME and SHAP explainability methods to interpret the predictions of our success prediction models.

\vspace{1mm} \noindent \textbf{Local Interpretable Model-agnostic Explanations (LIME)}.
Surrogate models are interpretable models (e.g., linear regression, decision tree) that are used to approximate the predictions of a black-box model. LIME \cite{lime} aims to train a local surrogate model to approximate the predictions of the black-box model for one specific instance (in our case, a student). Applying this method requires five steps: 1) select the student for which you want to explain the prediction, 2) generate new samples in the neighborhood of the selected student by slightly perturbing the feature values of the instance (i.e., generate new artificial students with slightly different feature values than the selected student), 3) weight each new synthetic student by the distance to the selected student (i.e. the higher the distance, the lower the weight), 4) train an interpretable surrogate model (e.g., a linear regression, decision tree) on the new perturbed feature values of the synthetic students, and 5) explain the prediction of the selected student by interpreting the surrogate model. The trained interpretable model therefore should be a good approximation of the black-box model locally. The outputs of LIME are the feature weights representing the feature influence on the final decision. The complexity of the surrogate model needs to be decided by the user. The default setting of LIME includes a maximum of $n=10$ features into the interpretable model. Therefore, if LIME is run with a linear regression and a maximum of $10$ features, LIME will return the regression weights of the optimal $10$ features. We decided to use this default setting for our evaluations in order to increase the reproducibility of our work. To ensure comparability between explainability methods, we scaled the obtained feature weights into importance scores to the interval $[0, 1]$, where $1$ indicates high feature importance.


\vspace{1mm} \noindent \textbf{SHapley Additive exPlanations (SHAP)}.
Inspired by game-theory's Shapley values \cite{shapley1997value} and LIME \cite{lime}, SHAP \cite{lundberg2017unified} is a popular method for explaining individual predictions. SHAP explains the prediction of an instance $s$ (in our case a student) by quantifying the contribution of each feature to the prediction. To determine a SHAP score for a feature, the algorithm has three main steps: 1) train a coalition of models on each possible subset of features (each feature individually, each combination of two features, each combination of three features, and so on), 2) identify marginal contributions of each feature (e.g. the difference in prediction of a model trained solely on $h_1$ and a model trained on features $h_1$ and $h_2$ to identify a marginal contribution of $h_2$), 3) accumulate these marginal contributions for each feature $h$ by computing a weighted sum. The computational and space complexity required for this algorithm is massive as it calls for $2^{|H|}$ models to be trained. 
To obtain comparable importance scores, we apply the same normalization transformation as LIME, taking the absolute values of the KernelSHAP feature attributions and scaling them to $[0, 1]$.

\vspace{1mm} \noindent \textbf{Counterfactual-based Confounder}.
\label{sec:confounder}
To validate our explainers with expert educators, we aim to showcase the results of the LIME and SHAP explainers and observe which explanations educators trust more across diverse course scenarios. However, without including a control explanation, we do not know if the LIME or SHAP explanations are really preferred, or just the best suboptimal option. To address this, we add a confounder explanation based on counterfactuals, using the contrastive explanation method \cite{dhurandhar2018explanations}. 
We purposely add inaccuracies into this approach by switching features from indicators of passing to failing (positive to negative importance scores).
In the study (detailed in Section \ref{sec:study}), we anonymize and shuffle LIME, SHAP, and the confounder and give all three explanations to expert educators.



\vspace{-2mm}
\subsection{Comparing Explanations}
\label{sec:quantitativecomparison}
The goal of our quantitative evaluation was to compare the feature importance scores across courses and methods and quantify the differences and similarities between them. Our choice of similarity metrics between methods is based on prior work. Specifically, \citet{Swamy2022} used \textit{Spearman's Rank-Order Correlation} to assess the agreement on the ordering of features (in terms of feature importance scores) between two different explainability methods $e_1$ and $e_2$. \textit{Spearman’s Rank-Order Correlation}, often referred to as \textit{Spearman's $\rho$} \cite{spearman1961proof}, identifies the rank correlation (statistical dependence between the rankings) between two variables. It is defined as the Pearson correlation coefficient between the rankings of two variables. We chose this metric to highlight the importance of feature ranking order in explanations.

In addition, we also use \textit{Jaccard Similarity} to assess the agreement on the top ranked features between $e_1$ and $e_2$ \cite{Swamy2022}. \textit{Jaccard Similarity} is used to identify the overlap between the top ten important features \cite{jaccard1901etude}. This method computes the cardinality of the intersection divided by the cardinality of the union of two feature sets. We chose this metric because of its approximation to the real-world setting of educational interventions based on explanations; an educator would only be able to intervene on a limited number of features. We refrained from comparing exact importance scores between methods, as LIME limits the complexity of the explanation (default $n=10$), resulting in sparser explanations (and thus generally higher importance scores). When comparing explainability methods across courses with different settings (which also entails different duration and different individual students), we average feature values across all weeks for each student, and then across all students to obtain an aggregated ranking.


\vspace{-2mm}
\subsection{Validating Explanations}
\label{sec:study}
To qualitatively validate the results of the explainers, we conduct a study with educational domain experts. To design the study, we conducted five pilot interviews with a mix of gender, experience, and educational research background. 

\vspace{1mm} \noindent \textbf{Participants}. The study was conducted with approval from the institutional review board for a target audience of experienced STEM professors (HREC 065-2022/27.09.2022). We interviewed $26$ educators for $30$ to $45$ minutes each, using the first five interviews as a pilot. The pilot interviewers provided feedback that we used to hone the visualizations of the XAI methods, the scenarios presented in the study, and the framing of the questions. We collected demographic information from our participants on age, gender, nationality, affiliated institution, teaching experience, and MOOC experience. The $21$ officially selected study participants are international; $33.33\%$ of our participants come from American universities, $23.81\%$ from Swiss universities, $14.28\%$ from Italian universities, $14.28\%$ from Middle Eastern universities, $9.52\%$ from Spanish universities, and $4.76\%$ from Australian universities. We have a high gender imbalance in our study, with $80.95\%$ of participants identifying as male. The mean age of the participants was 36.5 ($\sigma = 9$). All participants in the study were university-level educators with a technical background teaching science, engineering, data science, or computer science concepts. We chose these educators because the courses we chose to analyze in the study pertain to numerical analysis or functional programming taught at the university level, and we wanted to narrow the scope to experts in this specific educational sub-domain. $76.19\%$ of interviewees currently hold the job title of professor or lecturer; other participants are not officially recognized as researchers but taught or are currently teaching university courses. Several of the participants have experience with MOOCs and digital courses, with $33.33\%$ having been a participant in a MOOC, $57.14\%$ having taught a MOOC, and $28.57\%$ having worked with MOOC data. 

\vspace{1mm} \noindent \textbf{Procedure}. With our interviews, we examined two major educational dimensions in detail: setting (MOOC vs. flipped classroom) and active learning (high vs. low) using two separate scenarios. We chose to focus on these two aspects (out of our selected five) as they were the most discussed in related literature. Each participant only saw one scenario. 

In the first scenario (difference in setting), participants were told that they were a professor offering a flipped classroom course on programming. We provided a summary description of the course they teach, stating that it was offered by a European university to $218$ students for $10$ weeks, with both quizzes and videos, culminating in a final exam. We then mentioned that a fellow professor adapted their flipped classroom curriculum to a MOOC with the same topics and materials.
We explained to participants that neural network models had been trained on both courses, achieving high balanced accuracy (at least $90\%$) in terms of pass-fail performance prediction and that we had run three state-of-the art XAI methods (abstracted to M1, M2, and M3) to obtain feature importance scores. One of these methods was in fact an inaccurate confounder (see Section \ref{sec:confounder}), but we did not reveal this to the interview participants. 

Our XAI models selected $11$ features as particularly important for the first scenario. These features were constructed as the pair with the most positive importance score (contributing to student success) and most negative score (contributing to student failure) across each method for each course. The interviewees were asked (for each course) to rank two features they thought could indicate student success the most and two features they thought could indicate student failure the most; a feature like \textit{total-time-video} could perhaps positively indicate student success if the value was high and student failure if the value was low, so if they feel it is most important for both, they can select it in both sets. We used an interactive drag-drop interface (Google Jamboard) to help the participants conduct this ranking. The motivation for this was to build a prior for the interviewees before we show the explainability results. 

After the interviewees had completed their independent feature importance ranking for both courses, we showed the results for explainers M1, M2, and M3 as explainability importance plots (see the last section of Fig. \ref{fig:pipeline}). A large bar (either green or red) means that the XAI method found this feature important as either a passing or failing differentiator. We asked the participants two main questions: 1) Which explanation do they agree with the most and why? 2) Do they think the explainers agree on what is important (on a Likert scale from 1-5)? We repeat this analysis for both courses. We shuffled the XAI methods shown in the figures (i.e. LIME might be M1 for the first course and M3 for the second).

 
Finally, we showcase XAI insights across the difference in course setting. Using the feature importance distributions across all $110$ features, we found the two features that changed the most in importance positively across the difference (i.e. flipped to MOOC) and the two features that changed the most in a negative way. We also noted the most important weeks for these features (beginning, middle, end, or throughout the course). We asked interviewees two more questions: 3) Do they agree with these insights (on a Likert scale of 1-5)? 4) How could they use these insights actionably to transition a curriculum across the course setting difference? To evaluate these results, two authors of this paper coded the responses independently across five to six categories defined by looking at repetitive trends in interview transcripts. 

For the second scenario (active learning), the exact same interview procedure was followed, the only difference was related to the selected courses and the story. Participants were told that they were the professor of a numerical analysis MOOC on Coursera, offered worldwide and taught in French to $506$ students for 9 weeks, with a lot of videos and few quizzes (low active learning), culminating in a final exam. Their professor friend offered a similar MOOC in Geometry taught worldwide in French to $452$ students for $11$ weeks, with many more quizzes than videos (high active learning), culminating in a final exam. Eight features were selected as important for this selected course pair.




\section{Experimental Evaluation}
\label{sec:results}
We evaluated post-hoc explainability methods for models trained on five pairs of courses in order to compare feature importance across different course settings (Sec. \ref{sec:study-design}). Using these results, we first explored how feature importance varied across different explainers for one specific course. Then, we investigated the similarity of the explainability methods across each pair of courses using two distance metrics (Sec. \ref{sec:quantitativecomparison}). Finally, we presented real-world scenarios to educators to assess the validity and actionability of the explanations across courses and methods (Sec. \ref{sec:study}).

\subsection{Experimental Protocol}
We trained a separate BiLSTM model for each course. Hyperparameters were selected consistently with prior work \cite{Swamy2022}: two BiLSTM layers consisting of $64$, and $32$ units and one Dense layer consisting of $1$ unit with Sigmoid activation. As this work is not focused on improving model performance, we did not tune hyperparameters further. We split the data of each course into a training data set and a test data set, with $80\%$ and $20\%$ of students respectively. We performed a stratified train-test split over students’ pass-fail labels. For training and prediction, our models used the student log data collected for the full duration of the course. Balanced accuracy (BAC) was chosen as our primary evaluation metric because of the high-class imbalance of most of the selected courses.

\begin{figure*}
    \centering
     \includegraphics[trim={0 20 0 0},clip, width=\textwidth]{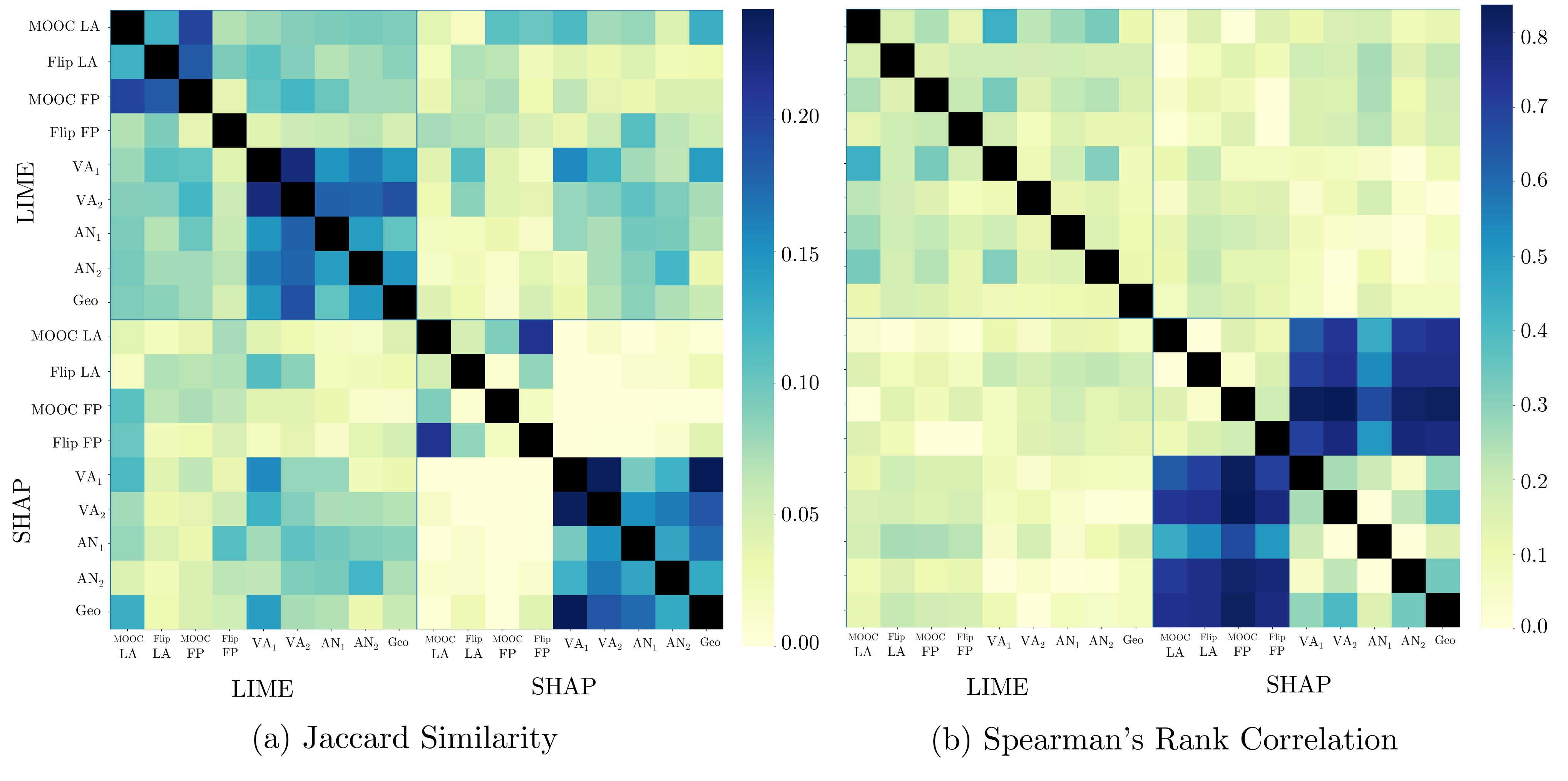}
     \caption{\textit{Jaccard Similarity} and \textit{Spearman’s Rank-Order Correlation} between XAI methods ordered by method.}
     \vspace{-3mm}
     \label{fig:heatmaps_big}
\end{figure*}

To analyze the explainers, we applied LIME and SHAP to compute the feature importances based on the model predictions for a specific instance. As it is not feasible to train each method for thousands of instances, we used a uniform sampling strategy to select $100$ representative students from each course, ensuring a balance between classes (pass-fail). We first extracted all failing students and ordered them according to the model's predicted probabilities. We then uniformly sampled $50$ failing students from this ordered interval. We repeated this exact same procedure to sample the $50$ passing students. This sampling procedure ensures that we include instances where the model is confident and wrong, where the model is unsure, and where the model is confident and correct.

\subsection{RQ1: Explanation Similarity}
In the first experiment, we compared the explanations of the instance-based methods within each course separately. We computed \textit{Jaccard Similarity} as well as \textit{Spearman’s Rank-Order Correlation} to assess distance between the top ten features and the overall feature rankings.

Fig. \ref{fig:heatmaps_big}a illustrates the \textit{Jaccard Similarity} between distributions of top ten important features over explainability methods. We observe clear patterns along the diagonal; this shows that there are similarities in the top 10 features among the same explainability method across different courses. In comparing courses across SHAP results (bottom left corner), we see high variance in the relationship with both the smallest scores in the graph and two strongly correlated sections. This implies that correlations across the SHAP method are heavily course dependent; the features SHAP finds important seem to depend on the similarity between courses. Examining SHAP and LIME, we note no strong correlations, indicating that these methods pick very different features for the same courses. We see an interestingly strong correlation using both explainability methods between $VA_1$ and $VA_2$. This might imply that the difference between French and English student populations are not as pervasive as expected.


Fig. \ref{fig:heatmaps_big}b uses \textit{Spearman's $\rho$} to compare explanations over all features. The patterns between the two metrics are not similar; examining significantly more than the top ten features and using ranks instead of set logic leads to a different conclusion. The one recognizable strong correlation is in comparing SHAP explanations over the four courses which differ with respect to the setting (FP and LA). In any other setting, no significant correlation exists, including in comparing LIME with itself. The distance between LIME and SHAP explanations is more evident using \textit{Spearman's $\rho$}.

\vspace{1mm} \noindent \textit{In summary,  we only observe agreement within method and not across explainability methods. Furthermore, the observed patterns for the two similarity metrics are very different, indicating that the choice of the similarity metric is essential.}



\subsection{RQ2: Contrasting Settings}
In a second analysis, we investigated whether the explainability methods could capture controlled differences across pairs of courses. While we performed comparisons for all five pairs of courses, we only display the two pairs presented to the educators in the user study (one pair for MOOC vs. flipped and one pair for active vs. passive learning). The results for the other three pairs are included in our Github repository.

\begin{figure}
    \centering
    \includegraphics[trim={0 20 0 0},clip, width=\textwidth]{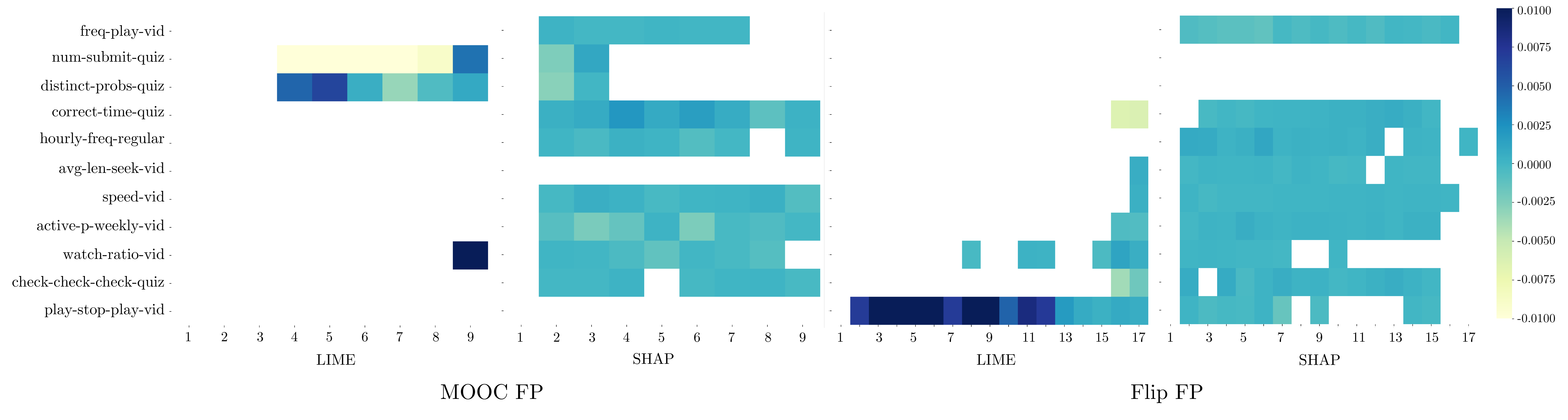}
    \caption{Flipped vs. MOOC Functional Programming course; showcasing absolute, normalized importance scores across weeks.}
    \label{fig:flip-mooc}
\end{figure}

\begin{figure}
    \centering
    \includegraphics[trim={0 15 0 0},clip, width=\textwidth]{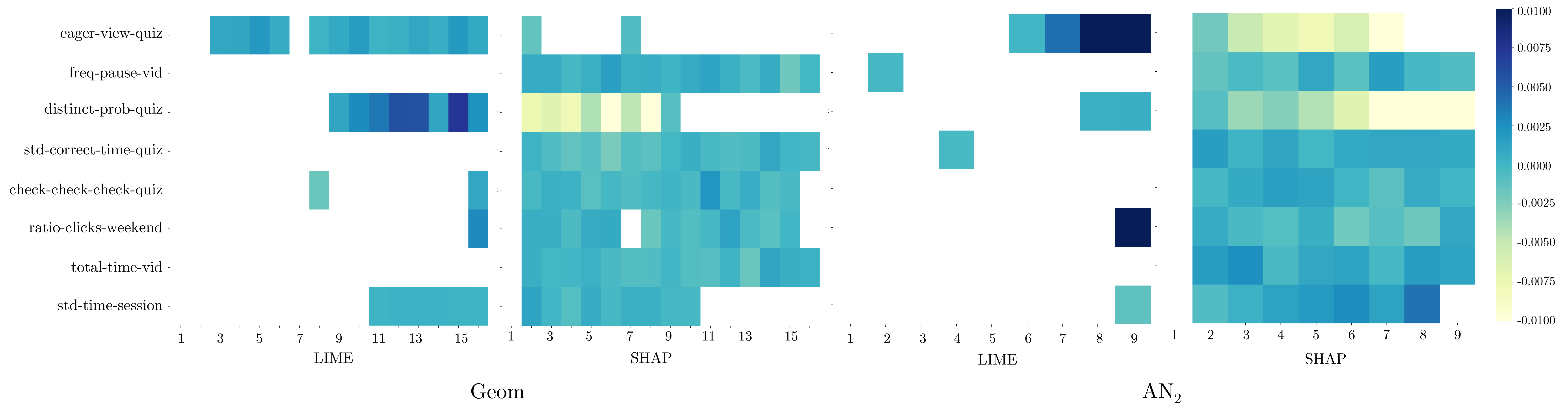}
    \caption{Low vs. High active learning: Geo and $AN_2$ course; showcasing absolute, normalized importance scores across weeks.}
    \label{fig:geom-an2}
\end{figure}

Fig. \ref{fig:flip-mooc} and Fig. \ref{fig:geom-an2} show heatmaps of feature importance scores aggregated across students for four courses. The selected features for each plot consist of the top positive feature and top negative feature for each method (LIME, SHAP, and the confounder) and course. This entails a potential maximum of $12$ features per pair; we obtain eight features for the active learning dimension (implying higher important feature overlap) and $11$ features for the setting dimension (low feature overlap). Feature importance scores below $|0.0001|$ are not displayed. We observe significant sparsity in LIME as opposed to SHAP, which could be useful in downstream interventions. The features are described in Table \ref{tab:features}. 

We examined more detailed relationships for each pair using the displayed importance scores as well as the top ten feature distributions for each method and course. At a high level, we observed in Fig. \ref{fig:flip-mooc} that LIME is far sparser than SHAP; SHAP tends to select substantial features with low weights, while LIME picks sparse features with high weights. Note that this is due to the default parameters of LIME, where the algorithm is restricted to ten features. In the MOOC setting, LIME chooses primarily problem features as important (notably \textit{num-submit-quiz} for failure and \textit{distinct-probs-quiz} for success), but in the flipped setting, LIME selects video features, especially \textit{play-stop-play-vid}, and focuses on the last weeks of the course. SHAP has a wide distribution, focusing on both video, quiz, and regularity features with a lack of emphasis on \textit{num-submit-quiz} and \textit{distinct-probs-quiz}, which are both indicative of student effort. In Fig. \ref{fig:geom-an2}, we see that \textit{distinct-probs-quiz} is regarded as important negative indicator of student performance for both courses by SHAP. However, in LIME, it is considered neutral or positive between both courses. LIME also places a positive weight on \textit{eager-view-quiz}, a proactivity feature, but SHAP considers it negatively. While the direction is different, we note that for the top polarized features, the absolute importance is recognized by both LIME and SHAP.

\vspace{1mm} \noindent \textit{In summary, a few learning science insights are supported and re-confirmed by the importance scores for both methods and course pairs. Video features are predictive across the board. Early week behavior is particularly important in MOOCs, likely due to dropout \cite{mooc-dropout}. In flipped classrooms and across differences in student population, specific problem patterns and timely engagement could be intuitive choices as course health metrics. Students who view or attempt elements early consistently (proactivity) are likely to pass. However, this section does not provide strongly generalizable insights due to a lack of scale.}

\subsection{RQ3: Expert Trust}
In our third analysis, to assess the validity of our explanations, we examined the results of $21$ interviews with non-pilot educators who ranked the four features they believed that differentiate the most passing students from failing students.

\textit{Active-participation-weekly-vid} was the most selected feature in the setting dimension, mostly as passing for Flip FP (six times) and equally as passing and failing for MOOC FP (four and five mentions respectively). \textit{Total-time-vid} was similarly often selected as an indicator for the high active learning setting, six times for passing and six times for failing. While video features were primarily considered as important by the educators, \textit{distinct-probs-quiz} was selected by experts seven time as a passing differentiator in the high active learning case (Geo).

After seeing the explanations from the three explainability methods (i.e., last section of Fig. \ref{fig:pipeline}), the educators chose the method they trusted most. In Flip FP, $80\%$ of participants trusted the SHAP explanation the most; the others were split equally between LIME and the confounder ($10\%$ each). For MOOC FP, the distribution was split almost evenly. $40\%$ of participants preferred the confounder, $30\%$ preferred LIME and SHAP each. Participants had the same average score for how much the methods agree with each other ($2.8$) with a slightly higher standard deviation for the MOOC course over the flipped course ($0.92$ and $0.79$ respectively). Only $2$ of $10$ participants chose the same method for both courses. A possible reason for this is that participants select the explanation per course that most confirms their a-priori beliefs (which often differ based on the educational aspect) instead of choosing consistently based on perceived explanation quality. This could also be interpreted as the participants' lack of clarity on whether the methods agree or disagree.

We also examined whether the features the participants selected at the beginning of the study correlated with their trusted explainer. In $26.49\%$ of cases, the features respectively selected by the participants as the most indicative of passing or failing were in agreement with the explainability method they chose. More specifically, they picked the pass-fail designation in agreement $38.6\%$ of the time in AN$_2$, $25\%$ of the time in Geo, $21\%$ of the time for Flip FP, and $20\%$ of the time for MOOC FP. Overall, study participants were better at picking passing features aligned with the explanation they trusted most over failing features ($29\%$ over $23\%$). However, none of these numbers surpass even $50\%$ accuracy, which means that the experts were not particularly accurate (with respect to their chosen explanation) about whether the features were more indicative of passing or failing students. We then hypothesized that even if the direction of important features was incorrect, perhaps the educators decided to choose important features aligned with their favored explanation nonetheless. $46.8\%$ of features the experts chose were in the top three features of the explanation they chose as important. Notably, for AN$_2$ (high active learning), educators selected features that were $68.2\%$ aligned with their chosen explanation. From this, we can conclude that educators chose features that are at least partially aligned with at least one explanation, validating the results of the explainers.

In the active learning subset, we observe much clearer preferences. For $AN_2$ with a high active learning score, a notable preference for the confounder ($45.5\%$) and SHAP ($54.5\%$) exists, with nobody trusting the LIME explanation. However, in the low active learning case (Geo), we see a strong preference for the LIME explanation ($81.8\%$), with the remaining $19.2\%$ preferring the confounder explanation. The difference between the preferences for similar courses is evident; experts do not agree on a gold standard explainer across different models, even with the same predictive method and feature sets. Only one participant chose the same method as most trustworthy for both courses. In determining the agreement of the XAI methods, the average rating was $2.09$ (on a Likert Scale of $5$) for $AN_2$ with standard deviation $0.94$ and $2.54$ for Geo with standard deviation $0.69$. This first result indicates that participants were very polarized on the $AN_2$ explanations, with some finding that the methods disagree extremely and some finding that they agree moderately; no participants in the low active learning setting said the explanations strongly agree with each other (a rating of 5).

\vspace{1mm} \noindent \textit{In summary, educators had diverse prior perceptions of what factors enable student success and failure; this led them to trust explanations that aligned with their beliefs (which led to significant disagreement across experts). Even more notably, experts were not consistent with their XAI method preferences across multiple courses; only three participants of the $21$ chose the same method for both courses in the pair.}

\subsection{RQ4: Actionable Course Design}
In our last analysis, we were interested in understanding how educators make course design decisions based on XAI insights. We found that $85.71\%$ educators were able to come up with at least one insight to improve course design. The two authors of this paper that conducted each interview independently coded each response with one or more categories depending on the suggestions given by the participant. We then computed Cohen's kappa on the binary outcome of tagging each response with each category, averaged over all categories. The inter-annotator agreement is high (a Cohen's kappa of $0.845$ for the active learning scenario and $0.84$ for the MOOC and flipped classroom scenario).

The large diversity of responses and the overwhelming ability of experts to find ideas based on XAI insights imply that XAI results are useful for improving course design and performing holistic analytics. To make XAI insights even more useful, we were often suggested to have more granularity or the ability to further drill down the XAI importance score based on feature distribution values. We also found that the inability to make clear linear causal statements was hard to grasp; we cannot say that a high amount of time watching videos is a contributor to student success, but we can say the amount of time watching videos is important for passing students (either positively or negatively correlated).

In the last part of the interview, we presented the important indicators of student success (and failure) as the features whose importance scores change the most positively (and negatively) across the educational facet. We also included a broad categorization of when in the course these features are important (beginning, middle, end, or throughout the course). For the setting dimension the presented success indicators were \textit{check-check-check-quiz} (end of course) and \textit{correct-time-quiz} (throughout the course). The failure indicators were \textit{attendance-rate} (end of course) and \textit{content-alignment} (middle of course). Based on these insights, $70\%$ of the interviewees suggested modifying the curriculum (e.g., reordering the course material). $55\%$ of answers suggested redistributing problems throughout the course. Other less popular suggestions mentioned by at least two participants involved pausing or shortening videos, including quizzes as part of videos, sending students reminders, and increasing cross-student interaction.

In the active learning setting, the important student success indicators were \textit{play-pause-load-vid} (start of course) and \textit{pause-speedchange-play-vid} (mid course). As failure indicators, we showed \textit{content anticipation} (end of course) and \textit{correct-time-quiz} (end of course). $45.5\%$ of responses focused on redistributing problems through the course; often the logic was that the time to correctly answer a quiz question was getting longer towards the course end because the quizzes were immediately diving into complex questions. Several educators proposed warming up to the difficult questions with easier initial review, or to spread question difficulty along weeks to discourage dropout towards the course end. $36.4\%$ of participants suggested shorter videos. $27.7\%$ suggested changing the structure (e.g., concepts ordering). Other suggestions included adding personalized student practice or sending deadline reminders.

The majority of participants across both scenarios agreed that the aggregated XAI indicators of student success and failure were logically feasible when comparing courses. We saw a feasibility agreement score of $4$ out of $5$ for the setting dimension and $3.81$ out of 5 for the active learning dimension, with a $0.82$ and $0.87$ standard deviation respectively. Hence, the comparison insights aggregated over all the explainers are more trustworthy than any explainer individually.

\vspace{1mm} \noindent \textit{In summary, expert educators found insights generated across contrastive course settings actionable. While the suggested actions differed across experts, over $85\%$ of educators were able to use insights to derive actions for improving course design.}

\section{Conclusion}
Over the past years, neural network models for education have gained in popularity. They are, however, still rarely used by educators. This is likely due to a lack of transparency, making actionable interventions impossible and decreasing educators' trust \cite{DBLP:conf/lak/NazaretskyCA22}. In this paper, we took a step towards tackling these explainer trustworthiness issues by 1) implementing explainability methods and student success prediction models over nine diverse course settings, 2) quantitatively validating explanations across course pairs that differ in one significant educational aspect, and 3) investigating educators trust in these explanations as well as their perceived usefulness for adapting course design. 

In a first analysis, we compared the different explainability methods across nine courses, including two flipped classrooms and seven MOOCs on EdX and Coursera. By using \textit{Jaccard Similarity} and \textit{Spearman's Rank-Correlation}, we found that the choice of the explainability method has a larger influence on the obtained feature importance score than the choice of the specific course, confirming the findings of prior work \cite{swamy22b}.

In a second step, we compared courses with contrastive settings. We analyzed the top ten features with the highest importance scores from LIME and SHAP. Our results demonstrate that 1) early week behavior is important in MOOCs, 2) interaction clickstreams are very important for flipped classrooms, 3) focusing on student behavior (i.e. lecture-watching delays) as a course health metric could be intuitive across differences in student population, and 4) the students who show anticipation (doing things early consistently) are generally very performant. While the two methods did not agree on the pass-fail predictive direction of the exact features, they agreed on the type of features important for course success. A limitation of this study is the scale, as we only examined five pairs of courses hosted by one university. Although we could not make strongly generalizable conclusions, we contribute to the corpora of educational science literature.

Thirdly, we conducted $21$ semi-structured expert interviews and five pilots to build trust in explainers. We identified that the educators were split amongst themselves across which methods they trust most, and many times even chose the confounder explanation which was designed to be misleading. From the interview feedback, many participants requested more concrete and granular insights and background about student demographics or past knowledge. We conclude that providing the averaged feature importance scores is not enough for educators to decide on concrete steps. It is possible that explainability methods are more effective on an individual scale (per student) than on a global scale (per course). We leave the further design of intuitive explanation visualizations, using more granular XAI insights, and measuring efficacy of instructor XAI-enabled suggestions to a future study. Note also that the selection of participants was biased by the networks of the authors of this paper (tending towards STEM professors); this is important in interpreting our results and the ease with which participants understood the technical background of the methods.

To conclude, XAI has a high potential in personalizing education, but the systematic differences between explainers is a significant barrier to educators trusting any one explanation. Future work is needed to make the explanations human interpretable. We observed that even without believing the explanations agree with each other, educators find that the provided insights can inspire ideas for improving course design and bolstering student performance.
\begin{acks}
This project was substantially co-financed by the Swiss State Secretariat
for Education, Research and Innovation (SERI). We thank Kate Kutsenok, Christian Giang, Jibril Frej, Peter Buhlman, Tanya Nazaretzky, and our 26 study participants for their generous time and support towards this project.
\end{acks}

\bibliographystyle{ACM-Reference-Format}
\bibliography{base}


\begin{thebibliography}{49}


\ifx \showCODEN    \undefined \def \showCODEN     #1{\unskip}     \fi
\ifx \showDOI      \undefined \def \showDOI       #1{#1}\fi
\ifx \showISBNx    \undefined \def \showISBNx     #1{\unskip}     \fi
\ifx \showISBNxiii \undefined \def \showISBNxiii  #1{\unskip}     \fi
\ifx \showISSN     \undefined \def \showISSN      #1{\unskip}     \fi
\ifx \showLCCN     \undefined \def \showLCCN      #1{\unskip}     \fi
\ifx \shownote     \undefined \def \shownote      #1{#1}          \fi
\ifx \showarticletitle \undefined \def \showarticletitle #1{#1}   \fi
\ifx \showURL      \undefined \def \showURL       {\relax}        \fi
\providecommand\bibfield[2]{#2}
\providecommand\bibinfo[2]{#2}
\providecommand\natexlab[1]{#1}
\providecommand\showeprint[2][]{arXiv:#2}

\bibitem[Conati et~al\mbox{.}()]%
        {conati2018ai}
\bibfield{author}{\bibinfo{person}{Cristina Conati}, \bibinfo{person}{Kaska
  Porayska-Pomsta}, {and} \bibinfo{person}{Manolis Mavrikis}.}
  \bibinfo{year}{2018}\natexlab{}.
\newblock \showarticletitle{\text{AI} in Education needs interpretable machine
  learning: Lessons from Open Learner Modelling}.
\newblock \bibinfo{journal}{\emph{International Conference on Machine
  Learning}} (\bibinfo{year}{2018}).
\newblock


\bibitem[Nazaretsky et~al\mbox{.}()]%
        {DBLP:conf/lak/NazaretskyCA22}
\bibfield{author}{\bibinfo{person}{Tanya Nazaretsky}, \bibinfo{person}{Mutlu
  Cukurova}, {and} \bibinfo{person}{Giora Alexandron}.}
  \bibinfo{year}{2022}\natexlab{}.
\newblock \showarticletitle{An Instrument for Measuring Teachers' Trust in
  AI-Based Educational Technology}.
\newblock \bibinfo{journal}{\emph{{Learning Analytics and Knowledge}}}
  (\bibinfo{year}{2022}).
\newblock


\bibitem[Hasib et~al\mbox{.}()]%
        {hasib2022lime}
\bibfield{author}{\bibinfo{person}{Khan~Md. Hasib}, \bibinfo{person}{Farhana
  Rahman}, \bibinfo{person}{Rashik Hasnat}, {and} \bibinfo{person}{Md.
  Golam~Rabiul Alam}.} \bibinfo{year}{2022}\natexlab{}.
\newblock \showarticletitle{A Machine Learning and Explainable \text{AI}
  Approach for Predicting Secondary School Student Performance}.
\newblock \bibinfo{journal}{\emph{IEEE Computing and Communication}}
  (\bibinfo{year}{2022}).
\newblock


\bibitem[Baranyi et~al\mbox{.}()]%
        {baranyi2020interpretable}
\bibfield{author}{\bibinfo{person}{M{\'a}t{\'e} Baranyi},
  \bibinfo{person}{Marcell Nagy}, {and} \bibinfo{person}{Roland Molontay}.}
  \bibinfo{year}{2020}\natexlab{}.
\newblock \showarticletitle{Interpretable deep learning for university dropout
  prediction}.
\newblock \bibinfo{journal}{\emph{Information Technology Education}}
  (\bibinfo{year}{2020}).
\newblock


\bibitem[Scheers and De~Laet()]%
        {scheers2021interactive}
\bibfield{author}{\bibinfo{person}{Hanne Scheers} {and} \bibinfo{person}{Tinne
  De~Laet}.} \bibinfo{year}{2021}\natexlab{}.
\newblock \showarticletitle{Interactive and Explainable Advising Dashboard
  Opens the Black Box of Student Success Prediction}.
\newblock \bibinfo{journal}{\emph{European Conference on Technology Enhanced
  Learning}} (\bibinfo{year}{2021}).
\newblock


\bibitem[Mu et~al\mbox{.}()]%
        {mu2020towards}
\bibfield{author}{\bibinfo{person}{Tong Mu}, \bibinfo{person}{Andrea Jetten},
  {and} \bibinfo{person}{Emma Brunskill}.} \bibinfo{year}{2020}\natexlab{}.
\newblock \showarticletitle{Towards Suggesting Actionable Interventions for
  Wheel-Spinning Students}.
\newblock \bibinfo{journal}{\emph{Educational Data Mining}}
  (\bibinfo{year}{2020}).
\newblock


\bibitem[Pei and Xing()]%
        {pei2021}
\bibfield{author}{\bibinfo{person}{Bo Pei} {and} \bibinfo{person}{Wanli Xing}.}
  \bibinfo{year}{2021}\natexlab{}.
\newblock \showarticletitle{An Interpretable Pipeline for Identifying At-Risk
  Students}.
\newblock \bibinfo{journal}{\emph{Journal of Education Computing Research}}
  (\bibinfo{year}{2021}).
\newblock


\bibitem[Swamy et~al\mbox{.}()]%
        {Swamy2022}
\bibfield{author}{\bibinfo{person}{Vinitra Swamy}, \bibinfo{person}{Bahar
  Radhmehr}, \bibinfo{person}{Natasa Krco}, \bibinfo{person}{Mirko Marras},
  {and} \bibinfo{person}{Tanja Kaser}.} \bibinfo{year}{2022}\natexlab{}.
\newblock \showarticletitle{Evaluating the Explainers: Black-Box Explainable
  Machine Learning for Student Success Prediction in {MOOCs}}.
\newblock \bibinfo{journal}{\emph{Educational Data Mining}}
  (\bibinfo{year}{2022}).
\newblock


\bibitem[Marras et~al\mbox{.}()]%
        {marras2021can}
\bibfield{author}{\bibinfo{person}{Mirko Marras}, \bibinfo{person}{Julien
  Tuang~Tu Vignoud}, {and} \bibinfo{person}{Tanja Kaser}.}
  \bibinfo{year}{2021}\natexlab{}.
\newblock \showarticletitle{Can Feature Predictive Power Generalize?
  Benchmarking Early Predictors of Student Success across Flipped and Online
  Courses}.
\newblock \bibinfo{journal}{\emph{Educational Data Mining}}
  (\bibinfo{year}{2021}).
\newblock


\bibitem[Swamy et~al\mbox{.}()]%
        {swamy22b}
\bibfield{author}{\bibinfo{person}{Vinitra Swamy}, \bibinfo{person}{Mirko
  Marras}, {and} \bibinfo{person}{Tanja K{\"{a}}ser}.}
  \bibinfo{year}{2022}\natexlab{}.
\newblock \showarticletitle{Meta Transfer Learning for Early Success Prediction
  in MOOCs}.
\newblock \bibinfo{journal}{\emph{Learning at Scale}} (\bibinfo{year}{2022}).
\newblock


\bibitem[Ribeiro et~al\mbox{.}()]%
        {lime}
\bibfield{author}{\bibinfo{person}{Marco~Tulio Ribeiro},
  \bibinfo{person}{Sameer Singh}, {and} \bibinfo{person}{Carlos Guestrin}.}
  \bibinfo{year}{2016}\natexlab{}.
\newblock \showarticletitle{Why Should {I} Trust You?: Explaining the
  Predictions of Any Classifier}.
\newblock \bibinfo{journal}{\emph{KDD}} (\bibinfo{year}{2016}).
\newblock


\bibitem[Lundberg and Lee()]%
        {shap}
\bibfield{author}{\bibinfo{person}{Scott~M. Lundberg} {and}
  \bibinfo{person}{Su-In Lee}.} \bibinfo{year}{2017}\natexlab{}.
\newblock \showarticletitle{A unified approach to interpreting model
  predictions}.
\newblock \bibinfo{journal}{\emph{Neural Information Processing Systems}}
  (\bibinfo{year}{2017}).
\newblock


\bibitem[Impey and Formanek()]%
        {impey2021moocs}
\bibfield{author}{\bibinfo{person}{Chris Impey} {and} \bibinfo{person}{Martin
  Formanek}.} \bibinfo{year}{2021}\natexlab{}.
\newblock \showarticletitle{{MOOCs} and 100 Days of COVID: Enrollment surges in
  massive open online astronomy classes during the coronavirus pandemic}.
\newblock \bibinfo{journal}{\emph{Social Sciences \& Humanities Open}}
  (\bibinfo{year}{2021}).
\newblock


\bibitem[Hirsto et~al\mbox{.}()]%
        {hirsto2019exploring}
\bibfield{author}{\bibinfo{person}{Laura Hirsto}, \bibinfo{person}{Sanna
  V{\"a}is{\"a}nen}, {and} \bibinfo{person}{Anni Arffman}.}
  \bibinfo{year}{2019}\natexlab{}.
\newblock \showarticletitle{Exploring students’ experiences of self-regulated
  learning during a large flipped classroom course in teacher education}.
\newblock \bibinfo{journal}{\emph{Journal of Learning, Teaching and Education
  Research}} (\bibinfo{year}{2019}).
\newblock


\bibitem[Ke and Kwak()]%
        {ke2013online}
\bibfield{author}{\bibinfo{person}{Fengfeng Ke} {and} \bibinfo{person}{Dean
  Kwak}.} \bibinfo{year}{2013}\natexlab{}.
\newblock \showarticletitle{Online learning across ethnicity and age: A study
  on learning interaction participation, perception, and learning
  satisfaction}.
\newblock \bibinfo{journal}{\emph{Computers \& Education}}
  (\bibinfo{year}{2013}).
\newblock


\bibitem[Ogan et~al\mbox{.}()]%
        {Ogan15}
\bibfield{author}{\bibinfo{person}{Amy Ogan}, \bibinfo{person}{Erin Walker},
  \bibinfo{person}{Ryan~S. Baker}, \bibinfo{person}{Ma. Mercedes~T. Rodrigo},
  \bibinfo{person}{Jose Carlo~A. Soriano}, {and}
  \bibinfo{person}{Maynor~Jimenez Castro}.} \bibinfo{year}{2015}\natexlab{}.
\newblock \showarticletitle{Towards Understanding How to Assess Help-Seeking
  Behavior Across Cultures}.
\newblock \bibinfo{journal}{\emph{Int. Journal Artif. Intell. Educ.}}
  (\bibinfo{year}{2015}).
\newblock


\bibitem[Rizvi et~al\mbox{.}()]%
        {Rizvi22}
\bibfield{author}{\bibinfo{person}{Saman Rizvi}, \bibinfo{person}{Bart
  Rienties}, \bibinfo{person}{Jekaterina Rogaten}, {and}
  \bibinfo{person}{René~F. Kizilcec}.} \bibinfo{year}{2022}\natexlab{}.
\newblock \showarticletitle{{Beyond one-size-fits-all in MOOCs: Variation in
  learning design and persistence of learners in different cultural and
  socioeconomic contexts}}.
\newblock \bibinfo{journal}{\emph{Computers in Human Behavior}}
  (\bibinfo{year}{2022}).
\newblock


\bibitem[Ruip{\'e}rez-Valiente et~al\mbox{.}()]%
        {ruiperez22}
\bibfield{author}{\bibinfo{person}{Jos{\'e}~A Ruip{\'e}rez-Valiente},
  \bibinfo{person}{Thomas Staubitz}, \bibinfo{person}{Matt Jenner},
  \bibinfo{person}{Sherif Halawa}, \bibinfo{person}{Jiayin Zhang},
  \bibinfo{person}{Ignacio Despujol}, \bibinfo{person}{Jorge
  Maldonado-Mahauad}, \bibinfo{person}{German Montoro},
  \bibinfo{person}{Melanie Peffer}, \bibinfo{person}{Tobias Rohloff},
  {et~al\mbox{.}}} \bibinfo{year}{2022}\natexlab{}.
\newblock \showarticletitle{Large scale analytics of global and regional MOOC
  providers: Differences in learners’ demographics, preferences, and
  perceptions}.
\newblock \bibinfo{journal}{\emph{Computers \& Education}}
  (\bibinfo{year}{2022}).
\newblock


\bibitem[Kizilcec et~al\mbox{.}()]%
        {Kiziclec13}
\bibfield{author}{\bibinfo{person}{Ren{\'{e}}~F. Kizilcec},
  \bibinfo{person}{Chris Piech}, {and} \bibinfo{person}{Emily Schneider}.}
  \bibinfo{year}{2013}\natexlab{}.
\newblock \showarticletitle{Deconstructing disengagement: analyzing learner
  subpopulations in {MOOCs}}.
\newblock \bibinfo{journal}{\emph{{Learning Analytics and Knowledge}}}
  (\bibinfo{year}{2013}).
\newblock


\bibitem[Gašević et~al\mbox{.}()]%
        {GASEVIC201668}
\bibfield{author}{\bibinfo{person}{Dragan Gašević}, \bibinfo{person}{Shane
  Dawson}, \bibinfo{person}{Tim Rogers}, {and} \bibinfo{person}{Danijela
  Gasevic}.} \bibinfo{year}{2016}\natexlab{}.
\newblock \showarticletitle{Learning analytics should not promote one size fits
  all: The effects of instructional conditions in predicting academic success}.
\newblock \bibinfo{journal}{\emph{The Internet and Higher Education}}
  (\bibinfo{year}{2016}).
\newblock


\bibitem[Jung et~al\mbox{.}()]%
        {JUNG2019377}
\bibfield{author}{\bibinfo{person}{Eulho Jung}, \bibinfo{person}{Dongho Kim},
  \bibinfo{person}{Meehyun Yoon}, \bibinfo{person}{Sanghoon Park}, {and}
  \bibinfo{person}{Barbara Oakley}.} \bibinfo{year}{2019}\natexlab{}.
\newblock \showarticletitle{The influence of instructional design on learner
  control, sense of achievement, and perceived effectiveness in a supersize
  MOOC course}.
\newblock \bibinfo{journal}{\emph{Computers \& Education}}
  (\bibinfo{year}{2019}).
\newblock


\bibitem[Yang et~al\mbox{.}()]%
        {Yang17}
\bibfield{author}{\bibinfo{person}{M. Yang}, \bibinfo{person}{Z. Shao},
  \bibinfo{person}{Q. Liu}, {and} \bibinfo{person}{C. Liu}.}
  \bibinfo{year}{2017}\natexlab{}.
\newblock \showarticletitle{{Understanding the quality factors that influence
  the continuance intention of students toward participation in {MOOCs}}}.
\newblock \bibinfo{journal}{\emph{Education Tech Research Dev}}
  (\bibinfo{year}{2017}).
\newblock


\bibitem[Cagiltay et~al\mbox{.}()]%
        {Cagiltay20}
\bibfield{author}{\bibinfo{person}{Nergiz~Ercil Cagiltay},
  \bibinfo{person}{Kursat Cagiltay}, {and} \bibinfo{person}{Berkan Celik}.}
  \bibinfo{year}{2020}\natexlab{}.
\newblock \showarticletitle{{An Analysis of Course Characteristics, Learner
  Characteristics, and Certification Rates in MITx {MOOCs}}}.
\newblock \bibinfo{journal}{\emph{The International Review of Research in Open
  and Distributed Learning}} (\bibinfo{year}{2020}).
\newblock


\bibitem[Hew()]%
        {Hew16}
\bibfield{author}{\bibinfo{person}{Khe~Foon Hew}.}
  \bibinfo{year}{2016}\natexlab{}.
\newblock \showarticletitle{{Promoting engagement in online courses: What
  strategies can we learn from three highly rated MOOCS}}.
\newblock \bibinfo{journal}{\emph{British Journal of Education Tech}}
  (\bibinfo{year}{2016}).
\newblock


\bibitem[Koedinger et~al\mbox{.}()]%
        {Koedinger15}
\bibfield{author}{\bibinfo{person}{Kenneth~R. Koedinger},
  \bibinfo{person}{Jihee Kim}, \bibinfo{person}{Julianna~Zhuxin Jia},
  \bibinfo{person}{Elizabeth~A. McLaughlin}, {and} \bibinfo{person}{Norman~L.
  Bier}.} \bibinfo{year}{2015}\natexlab{}.
\newblock \showarticletitle{{Learning is Not a Spectator Sport: Doing is Better
  than Watching for Learning from a MOOC}}.
\newblock \bibinfo{journal}{\emph{Learning at Scale}} (\bibinfo{year}{2015}).
\newblock


\bibitem[Mejia{-}Domenzain et~al\mbox{.}()]%
        {mejia22}
\bibfield{author}{\bibinfo{person}{Paola Mejia{-}Domenzain},
  \bibinfo{person}{Mirko Marras}, \bibinfo{person}{Christian Giang}, {and}
  \bibinfo{person}{Tanja K{\"{a}}ser}.} \bibinfo{year}{2022}\natexlab{}.
\newblock \showarticletitle{Identifying and Comparing Multi-dimensional Student
  Profiles Across Flipped Classrooms}.
\newblock \bibinfo{journal}{\emph{{Artificial Intelligence in Education}}}
  (\bibinfo{year}{2022}).
\newblock


\bibitem[Hardebolle et~al\mbox{.}()]%
        {hardebolle2022gender}
\bibfield{author}{\bibinfo{person}{C{\'e}cile Hardebolle},
  \bibinfo{person}{Himanshu Verma}, \bibinfo{person}{Roland Tormey}, {and}
  \bibinfo{person}{Simone Deparis}.} \bibinfo{year}{2022}\natexlab{}.
\newblock \showarticletitle{Gender, prior knowledge, and the impact of a
  flipped linear algebra course for engineers over multiple years}.
\newblock \bibinfo{journal}{\emph{Journal of Engineering Education}}
  (\bibinfo{year}{2022}).
\newblock


\bibitem[Gardner and Brooks()]%
        {gardner2018student}
\bibfield{author}{\bibinfo{person}{Josh Gardner} {and}
  \bibinfo{person}{Christopher Brooks}.} \bibinfo{year}{2018}\natexlab{}.
\newblock \showarticletitle{Student success prediction in \text{MOOCs}}.
\newblock \bibinfo{journal}{\emph{User Modeling and User-Adapted Interaction}}
  (\bibinfo{year}{2018}).
\newblock


\bibitem[Lee et~al\mbox{.}()]%
        {lee2022affects}
\bibfield{author}{\bibinfo{person}{Jihyun Lee}, \bibinfo{person}{Taejung Park},
  {and} \bibinfo{person}{Robert~Otto Davis}.} \bibinfo{year}{2022}\natexlab{}.
\newblock \showarticletitle{What affects learner engagement in flipped learning
  and what predicts its outcomes?}
\newblock \bibinfo{journal}{\emph{British Journal of Education Tech}}
  (\bibinfo{year}{2022}).
\newblock


\bibitem[Akpinar et~al\mbox{.}()]%
        {DBLP:conf/edm/AkpinarRA20}
\bibfield{author}{\bibinfo{person}{Nil{-}Jana Akpinar},
  \bibinfo{person}{Aaditya Ramdas}, {and} \bibinfo{person}{Umut Acar}.}
  \bibinfo{year}{2020}\natexlab{}.
\newblock \showarticletitle{Analyzing Student Strategies In Blended Courses
  Using Clickstream Data}.
\newblock \bibinfo{journal}{\emph{{Educational Data Mining}}}
  (\bibinfo{year}{2020}).
\newblock


\bibitem[Boroujeni et~al\mbox{.}()]%
        {boroujeni2016quantify}
\bibfield{author}{\bibinfo{person}{Mina~Shirvani Boroujeni},
  \bibinfo{person}{Kshitij Sharma}, \bibinfo{person}{{\L}ukasz Kidzi{\'n}ski},
  \bibinfo{person}{Lorenzo Lucignano}, {and} \bibinfo{person}{Pierre
  Dillenbourg}.} \bibinfo{year}{2016}\natexlab{}.
\newblock \showarticletitle{How to quantify student’s regularity?}
\newblock \bibinfo{journal}{\emph{European Conference on Technology Enhanced
  Learning}} (\bibinfo{year}{2016}).
\newblock


\bibitem[Chen and Cui()]%
        {chen2020utilizing}
\bibfield{author}{\bibinfo{person}{Fu Chen} {and} \bibinfo{person}{Ying Cui}.}
  \bibinfo{year}{2020}\natexlab{}.
\newblock \showarticletitle{Utilizing Student Time Series Behaviour in Learning
  Management Systems for Early Prediction of Course Performance}.
\newblock \bibinfo{journal}{\emph{Journal of Learning Analytics}}
  (\bibinfo{year}{2020}).
\newblock


\bibitem[Lall{\'e} and Conati()]%
        {lalle2020data}
\bibfield{author}{\bibinfo{person}{S{\'e}bastien Lall{\'e}} {and}
  \bibinfo{person}{Cristina Conati}.} \bibinfo{year}{2020}\natexlab{}.
\newblock \showarticletitle{A data-driven student model to provide adaptive
  support during video watching across \text{MOOCs}}.
\newblock \bibinfo{journal}{\emph{Artificial Intelligence in Education}}
  (\bibinfo{year}{2020}).
\newblock


\bibitem[Lemay and Doleck()]%
        {DBLP:journals/eait/LemayD20}
\bibfield{author}{\bibinfo{person}{David~John Lemay} {and}
  \bibinfo{person}{Tenzin Doleck}.} \bibinfo{year}{2020}\natexlab{}.
\newblock \showarticletitle{Grade prediction of weekly assignments in {MOOCs:}
  mining video-viewing behavior}.
\newblock \bibinfo{journal}{\emph{Education and Information Technologies}}
  (\bibinfo{year}{2020}).
\newblock


\bibitem[Mbouzao et~al\mbox{.}()]%
        {DBLP:conf/aied/MbouzaoDS20}
\bibfield{author}{\bibinfo{person}{Boniface Mbouzao},
  \bibinfo{person}{Michel~C. Desmarais}, {and} \bibinfo{person}{Ian Shrier}.}
  \bibinfo{year}{2020}\natexlab{}.
\newblock \showarticletitle{Early Prediction of Success in {MOOC} from Video
  Interaction Features}.
\newblock \bibinfo{journal}{\emph{Artificial Intelligence in Education}}
  (\bibinfo{year}{2020}).
\newblock


\bibitem[Mubarak et~al\mbox{.}()]%
        {DBLP:journals/eait/MubarakCA21}
\bibfield{author}{\bibinfo{person}{Ahmed~Ali Mubarak}, \bibinfo{person}{Han
  Cao}, {and} \bibinfo{person}{Salah A.~M. Ahmed}.}
  \bibinfo{year}{2021}\natexlab{}.
\newblock \showarticletitle{Predictive learning analytics using deep learning
  model in \text{MOOC}s' courses videos}.
\newblock \bibinfo{journal}{\emph{Education and Information Technologies}}
  (\bibinfo{year}{2021}).
\newblock


\bibitem[Wan et~al\mbox{.}()]%
        {DBLP:journals/tlt/WanLYG19}
\bibfield{author}{\bibinfo{person}{Han Wan}, \bibinfo{person}{Kangxu Liu},
  \bibinfo{person}{Qiaoye Yu}, {and} \bibinfo{person}{Xiaopeng Gao}.}
  \bibinfo{year}{2019}\natexlab{}.
\newblock \showarticletitle{Pedagogical Intervention Practices: Improving
  Learning Engagement Based on Early Prediction}.
\newblock \bibinfo{journal}{\emph{{IEEE} Transactions on Learning
  Technologies}} (\bibinfo{year}{2019}).
\newblock


\bibitem[Broadbent and Poon()]%
        {broadbent2015self}
\bibfield{author}{\bibinfo{person}{Journal Broadbent} {and}
  \bibinfo{person}{W.L. Poon}.} \bibinfo{year}{2015}\natexlab{}.
\newblock \showarticletitle{Self-regulated learning strategies \& academic
  achievement in online higher education learning environments: A systematic
  review}.
\newblock \bibinfo{journal}{\emph{Internet Higher Education}}
  (\bibinfo{year}{2015}).
\newblock


\bibitem[He et~al\mbox{.}()]%
        {he2018measuring}
\bibfield{author}{\bibinfo{person}{Huan He}, \bibinfo{person}{Qinghua Zheng},
  \bibinfo{person}{Bo Dong}, {and} \bibinfo{person}{Hongchao Yu}.}
  \bibinfo{year}{2018}\natexlab{}.
\newblock \showarticletitle{Measuring Student's Utilization of Video Resources
  and Its Effect on Academic Performance}.
\newblock \bibinfo{journal}{\emph{International Conference on Advanced Learning
  Technologies}} (\bibinfo{year}{2018}).
\newblock


\bibitem[Cho and Shen()]%
        {cho2013self}
\bibfield{author}{\bibinfo{person}{Moon-Heum Cho} {and} \bibinfo{person}{Demei
  Shen}.} \bibinfo{year}{2013}\natexlab{}.
\newblock \showarticletitle{Self-regulation in online learning}.
\newblock \bibinfo{journal}{\emph{Distance Education.}} (\bibinfo{year}{2013}).
\newblock


\bibitem[Jovanovic et~al\mbox{.}()]%
        {DBLP:journals/ce/JovanovicMGDP19}
\bibfield{author}{\bibinfo{person}{Jelena Jovanovic}, \bibinfo{person}{Negin
  Mirriahi}, \bibinfo{person}{Dragan Ga{\v{s}}evi{\'c}}, \bibinfo{person}{Shane
  Dawson}, {and} \bibinfo{person}{Abelardo Pardo}.}
  \bibinfo{year}{2019}\natexlab{}.
\newblock \showarticletitle{Predictive power of regularity of pre-class
  activities in a flipped classroom}.
\newblock \bibinfo{journal}{\emph{Computers \& Education}}
  (\bibinfo{year}{2019}).
\newblock


\bibitem[Geertshuis et~al\mbox{.}()]%
        {geertshuis2014preparing}
\bibfield{author}{\bibinfo{person}{Susan Geertshuis}, \bibinfo{person}{Moon
  Jung}, {and} \bibinfo{person}{Helena Cooper-Thomas}.}
  \bibinfo{year}{2014}\natexlab{}.
\newblock \showarticletitle{Preparing Students for Higher Education: The Role
  of Proactivity.}
\newblock \bibinfo{journal}{\emph{International Journal of Teaching and
  Learning in Higher Education}} (\bibinfo{year}{2014}).
\newblock


\bibitem[Biard et~al\mbox{.}()]%
        {biard2018effects}
\bibfield{author}{\bibinfo{person}{Nicolas Biard}, \bibinfo{person}{Salom{\'e}
  Cojean}, {and} \bibinfo{person}{Eric Jamet}.}
  \bibinfo{year}{2018}\natexlab{}.
\newblock \showarticletitle{Effects of segmentation and pacing on procedural
  learning by video}.
\newblock \bibinfo{journal}{\emph{Computers in Human Behavior}}
  (\bibinfo{year}{2018}).
\newblock


\bibitem[Shapley()]%
        {shapley1997value}
\bibfield{author}{\bibinfo{person}{Lloyd~S Shapley}.}
  \bibinfo{year}{1997}\natexlab{}.
\newblock \showarticletitle{A value for n-person games}.
\newblock \bibinfo{journal}{\emph{Classics in game theory}}
  (\bibinfo{year}{1997}).
\newblock


\bibitem[Lundberg and Lee()]%
        {lundberg2017unified}
\bibfield{author}{\bibinfo{person}{Scott~M Lundberg} {and}
  \bibinfo{person}{Su-In Lee}.} \bibinfo{year}{2017}\natexlab{}.
\newblock \showarticletitle{A unified approach to interpreting model
  predictions}.
\newblock \bibinfo{journal}{\emph{Neural Information Processing Systems}}
  (\bibinfo{year}{2017}).
\newblock


\bibitem[Dhurandhar et~al\mbox{.}()]%
        {dhurandhar2018explanations}
\bibfield{author}{\bibinfo{person}{Amit Dhurandhar}, \bibinfo{person}{Pin-Yu
  Chen}, \bibinfo{person}{Ronny Luss}, \bibinfo{person}{Chun-Chen Tu},
  \bibinfo{person}{Paishun Ting}, \bibinfo{person}{Karthikeyan Shanmugam},
  {and} \bibinfo{person}{Payel Das}.} \bibinfo{year}{2018}\natexlab{}.
\newblock \showarticletitle{Explanations based on the missing: Towards
  contrastive explanations with pertinent negatives}.
\newblock \bibinfo{journal}{\emph{Neural Information Processing Systems}}
  (\bibinfo{year}{2018}).
\newblock


\bibitem[Spearman()]%
        {spearman1961proof}
\bibfield{author}{\bibinfo{person}{Charles Spearman}.}
  \bibinfo{year}{1961}\natexlab{}.
\newblock \showarticletitle{The proof and measurement of association between
  two things}.
\newblock  (\bibinfo{year}{1961}).
\newblock


\bibitem[Jaccard()]%
        {jaccard1901etude}
\bibfield{author}{\bibinfo{person}{Paul Jaccard}.}
  \bibinfo{year}{1901}\natexlab{}.
\newblock \showarticletitle{{Comparative study of the floral distribution in a
  portion of the Alps and the Jura}}.
\newblock  (\bibinfo{year}{1901}).
\newblock


\bibitem[Jordan()]%
        {mooc-dropout}
\bibfield{author}{\bibinfo{person}{Katy Jordan}.}
  \bibinfo{year}{2014}\natexlab{}.
\newblock \showarticletitle{Initial Trends in Enrollment and Completion of
  Massive Open Online Courses}.
\newblock \bibinfo{journal}{\emph{Research in Open and Distributed Learning}}
  (\bibinfo{year}{2014}).
\newblock


\end{thebibliography}

\end{document}